\documentclass[aps,prl,twocolumn,superscriptaddress,showpacs]{revtex4-1}

\usepackage{graphicx}

\usepackage{color}
\newcommand{\vlowk}{V_{{\rm low}\,k}}

\newcommand{\kev}{\, \text{keV}}

\newcommand{\bbz}{0 \nu\beta\beta \textup{ decay}}
\newcommand{\bbt}{2 \nu\beta\beta \textup{ decay}}
\newcommand{\bb}{\beta\beta \textup{ decay}}
\newcommand{\qbb}{Q_{\beta\beta}}

\begin{document}


\title{First Direct Double-Beta Decay $Q$-value Measurement of $^{82}$Se in Support of Understanding the Nature of the Neutrino}


\author{David L. Lincoln}
\email[]{lincoln@nscl.msu.edu}
\affiliation{Department of Physics and Astronomy, Michigan State University, East Lansing, Michigan, 48824, USA}
\affiliation{National Superconducting Cyclotron Laboratory, East Lansing, Michigan, 48824, USA}
\author{Jason D.\ Holt}
\affiliation{Department of Physics and Astronomy, University of Tennessee, Knoxville, TN, 37996, USA}
\affiliation{Physics Division, Oak Ridge National Laboratory, Oak Ridge, TN, 37831, USA}
\affiliation{Institut f\"ur Kernphysik, Technische Universit\"at Darmstadt, 64289 Darmstadt, Germany}
\affiliation{ExtreMe Matter Institute EMMI, GSI Helmholtzzentrum f\"ur Schwerionenforschung GmbH, 64291 Darmstadt, Germany}
\author{Georg Bollen}
\affiliation{Department of Physics and Astronomy, Michigan State University, East Lansing, Michigan, 48824, USA}
\affiliation{National Superconducting Cyclotron Laboratory, East Lansing, Michigan, 48824, USA}
\author{Maxime Brodeur}
\affiliation{National Superconducting Cyclotron Laboratory, East Lansing, Michigan, 48824, USA}
\author{Scott Bustabad}
\affiliation{Department of Physics and Astronomy, Michigan State University, East Lansing, Michigan, 48824, USA}
\affiliation{National Superconducting Cyclotron Laboratory, East Lansing, Michigan, 48824, USA}
\author{Jonathan Engel}
\affiliation{Department of Physics and Astronomy, University of North Carolina, Chapel Hill, North Carolina, 27599, USA}
\author{Samuel J. Novario}
\affiliation{Department of Physics and Astronomy, Michigan State University, East Lansing, Michigan, 48824, USA}
\affiliation{National Superconducting Cyclotron Laboratory, East Lansing, Michigan, 48824, USA}
\author{Matthew Redshaw}
\affiliation{National Superconducting Cyclotron Laboratory, East Lansing, Michigan, 48824, USA}
\affiliation{Department of Physics, Central Michigan University, Mount Pleasant, Michigan, 48859, USA}
\author{Ryan Ringle}
\affiliation{National Superconducting Cyclotron Laboratory, East Lansing, Michigan, 48824, USA}
\author{Stefan Schwarz}
\affiliation{National Superconducting Cyclotron Laboratory, East Lansing, Michigan, 48824, USA}


\date{\today}

\begin{abstract}

In anticipation of results from current and future double-beta decay
studies, we report a measurement resulting in a $^{82}$Se double-beta decay
$Q$-value of 2997.9(3)$\kev$, an order of magnitude more precise than the
currently accepted value. We also present preliminary results of a calculation
of the $^{82}$Se neutrinoless double-beta decay nuclear matrix element that
corrects in part for the small size of the shell model single-particle space.
The results of this work are important for designing next generation 
double-beta decay experiments and for the theoretical interpretations of their
observations.

\end{abstract}


\maketitle



The results of recent neutrino oscillation experiments indicate that the mass
of the neutrino is non-zero \cite{RefAbe2008,RefAdamson2008,RefAharmim}. The
mass hierarchy and the absolute mass scale of the neutrino, however, are
unknown. Furthermore, the nature of the neutrino is also unknown; is it a
Dirac or Majorana particle, i.e. is the neutrino its own antiparticle? The
only known practical method for determining the nature of the neutrino is
through neutrinoless double-beta decay ($\bbz$) measurements
\cite{RefAvignone2008}. Interest in double-beta decay ($\bb$) has been
increasing since the laboratory verification of the weak, but allowed,
two-neutrino double-beta decay ($\bbt$) of $^{82}$Se \cite{RefElliot1987}.
Including laboratory, geochemical, and radiochemical experiments, twelve
isotopes have been observed to undergo $\bbt$: $^{48}$Ca, $^{76}$Ge,
$^{82}$Se, $^{96}$Zr, $^{100}$Mo, $^{116}$Cd, $^{128}$Te, $^{130}$Te,
$^{136}$Xe, $^{150}$Nd, $^{238}$U, and $^{130}$Ba \cite{RefBarabash2010,
RefAckerman2011}. With the exception of the controversial claim in Ref.\
\cite{RefKlapdor2001}, $\bbz$ has yet to be observed. If experiments succeed
in observing $\bbz$, we would have evidence that the neutrino is a Majorana
particle and that conservation of total lepton number is violated --- a
situation forbidden by the standard model of particle physics.

An extensive campaign is currently underway to develop next-generation
experiments to detect $\bbz$ in a number of candidate isotopes (see Ref.\
\cite{RefBarabash2011} for a current review of planned experiments). One such
experiment, SuperNEMO, is expected to provide an increase in sensitivity of
three orders of magnitude over its predecessor, NEMO-III, and is projected to
reach a half-life sensitivity at the 90\% confidence level of 1-2 x 10$^{26}$
years by observing 100-200 kg of $^{82}$Se for five years
\cite{RefBarabash2011, RefSaakyan2009}.

The defining observable of $\bbz$ is a single peak in the electron sum-energy
spectrum at the $\beta\beta$ decay $Q$-value, $\qbb$. Hence, it is crucial to
have an accurate determination of $\qbb$. The $Q$-value is also a key
parameter required to determine the phase space factor (PSF) of the decay. The
effective Majorana neutrino mass, together with the corresponding PSF and
nuclear matrix element (NME) for a $\bbz$ candidate provide the necessary
information to determine the $\bbz$ half-life, which is given by:
\begin{equation}
(T^{0\nu}_{1/2})^{-1} = G_{0\nu}({Q_{\beta\beta}}^{5},Z)\vert M_{0\nu}\vert
^{2}(\langle\it{m_{\beta\beta}}\rangle/m_e) ^{2},
\end{equation}
where $\it{M}$ is the relevant NME, $\langle\it{m_{\beta\beta}}\rangle$ is the
effective Majorana mass of the neutrino, $m_e$ is the mass of the electron,
and $\it{G_{0\nu}}$ is the PSF for the $\bbz$, which is a function of
$\it{Q{_{\beta\beta}}}$$^{5}$ and the nuclear charge, $\it{Z}$. Thus, to
obtain an accurate estimation of the half-life sensitivity needed to detect a
given $\langle\it{m_{\beta\beta}}\rangle$, or conversely, to determine
$\langle\it{m_{\beta\beta}}\rangle$ if the half-life is measured, the NME and
especially the $Q$-value need to be known to high precision.

Of all the $\bbz$ candidates currently employed in experiments, $^{82}$Se is
the only one whose $Q$-value has not been measured directly through
high-precision Penning trap mass spectrometry (PTMS). PTMS has proven itself
to be the most precise and accurate method for determining atomic masses and
therefore, $Q$-values \cite{Lunney2003}. In some cases $\qbb$-values
determined prior to direct Penning trap measurements have been found to be off
by more than ten keV \cite{RefFink2012, RefRedshaw2012}. Furthermore, careful
calculations of the NME for $^{82}$Se $\bbz$ \cite{SM,SM2,QRPA,IBM,GCM} differ
from one another by more than a factor of two. In this letter, in anticipation
of SuperNEMO and other possible experiments with $^{82}$Se, we present the
results of a direct measurement of $\qbb$ with the Low-Energy Beam and Ion
Trap (LEBIT) PTMS facility and results to improve shell model calculations of
the NME.


The direct $\qbb$ measurement for $^{82}$Se was carried out at the National
Superconducting Cyclotron Laboratory (NSCL) using LEBIT, a Penning trap mass
spectrometer facility for high precision mass measurements on isotopes
produced via projectile fragmentation \cite{RefBollen2006}. We used the plasma
ion source of the LEBIT facility to simultaneously produce ions of $^{82}$Se
and of the $\bb$ daughter, $^{82}$Kr. The source was equipped with a ceramic
charge holder and was filled with $\sim$ 200 mg of granulated selenium which
was vaporized by the heat from the ion source filament. A helium support gas
for the source was mixed with a small amount of krypton to obtain a similar
beam current of $^{82}$Kr$^{+}$ to $^{82}$Se$^{+}$ within a factor of three.
The extracted ion beam was guided through a radio-frequency quadrupole (RFQ)
mass filter to suppress the strong accompanying helium current before
injection into a cryogenic RFQ beam cooler and buncher for the creation of short
low-emittance ion bunches that are sent to the LEBIT 9.4 T Penning trap
\cite{RefSchwarz2003}. On their path the ions were purified further by using a
time-of-flight mass separation scheme \cite{RefSchury2006}, allowing
only ion species of the same mass-to-charge ratio to be dynamically captured
in the trap.

At LEBIT, the cyclotron frequency, $\it{\nu_{c} = qB/2\pi m}$, of an ion with
mass-to-charge ratio, $\it{m/q}$, in a magnetic field, $\it{B}$ is measured
using the Time-of-Flight (TOF) ion-cyclotron resonance detection technique
\cite{RefKonig1995,RefBollen1990}. First, isobaric contaminants are removed
from the Penning trap by driving them to large radial orbits with a radio
frequency (RF) azimuthal dipole field. To measure $\it{\nu_{c}}$, the trapped
ions are exposed to an azimuthal quadrupole RF field at a frequency
$\it{\nu_{RF}}$ near their cyclotron frequency with the appropriate RF amplitude
and excitation time \cite{RefBollen1990,RefKonig1995}. After ejection from the
trap, the ions travel through the inhomogeneous section of the magnetic field,
where the energy of the ions' radial motion is transferred to the axial
direction \cite{RefGraff1980}, and are detected on a micro-channel plate (MCP)
detector. In resonance, i.e., $\it{\nu_{RF} = \nu_{c}}$, the energy pickup of
the ions' radial motion is maximized and their TOF to the MCP is
minimized \cite{RefKonig1995}. For a cyclotron frequency determination, this
cycle of trapping, excitation, ejection, and TOF measurement is repeated for
different frequencies. This leads to cyclotron resonance curves, as shown in Figure
1, with a centroid at $\it{\nu_{c}}$.

The measurement process for the determination of $\qbb$($^{82}$Se) consisted
of alternating cyclotron frequency measurements of $^{82}$Kr$^{+}$ and
$^{82}$Se$^{+}$. These measurements were performed in a series of four runs.
The first run consisted of measurements using a 500 ms excitation time. For
increased precision, we utilized a 750 ms excitation time for the final three
runs. Each TOF resonance was the average of 25 to 40 scans over the respective
frequency range with 41 trapping cycles per scan. During the measurement
process the number of ions recorded by the MCP was limited to an average of 2
ions per trapping cycle, corresponding to $<$ 7 ions in the trap at a time
(assuming 30”\%” detector efficiency). This was done to limit the number of
contaminant ions produced via charge-exchange reactions. Each resonance
consisted of $\sim$ 500 to 3000 detected ions, depending on the number of
scans per resonance and the beam current from the ion source. To determine
$\it{\nu_{c}}$, each resonance was fitted using the theoretical line shape
described in Ref.\ \cite{RefKonig1995}. The average uncertainty in
$\it{\nu_{c}}$ for each resonance was $\sim$ 30 ppb (parts per billion).

\begin{figure}[t]
\begin{center}
\includegraphics[scale=0.308,clip=]{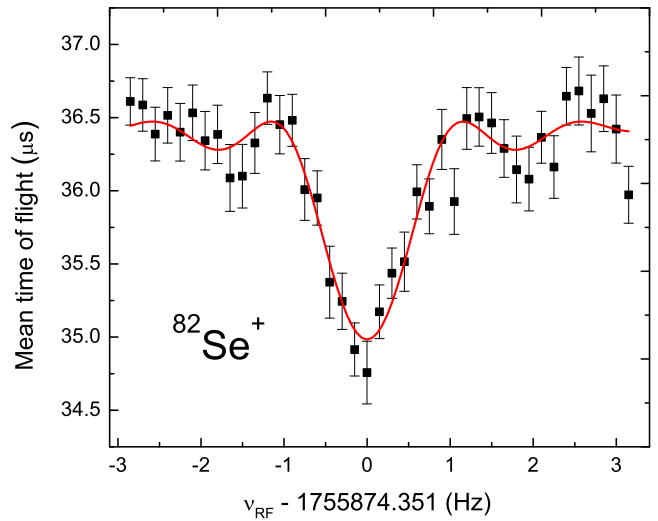}
\end{center}
\caption{(color online). Time-of-flight cyclotron resonance curve for
$^{82}$Se$^{+}$. An excitation time of $\it{T_{RF}}$ = 750 ms was used to
obtain a resolving power of 2 x 10$^6$. The fit of the theoretical line shape
to the data is represented by the solid line.\label{Fig:2}}
\end{figure}

For the frequency ratio determination of $^{82}$Kr$^{+}$ to $^{82}$Se$^{+}$,
drifts in the magnetic field were taken into account by linearly interpolating
two cyclotron frequency measurements of $^{82}$Kr$^{+}$ bracketing a
$^{82}$Se$^{+}$ measurement to obtain $\it{\nu^{int}_{c}}$($^{82}$Kr$^{+}$).
This interpolated cyclotron frequency was used to obtain the frequency ratio
$\it{R=\nu^{int}_{c}}$($^{82}$Kr$^{+}$)$\it/\nu_c$($^{82}$Se$^{+}$). The
values obtained from a total of 110 ratio determinations and their weighted
average are shown in Figure 2. The difference to the reference ratio,
$\it{R_{\mathrm {AME2003}}=[{m}}$($^{82}$Kr) $\it-$ $\it m_e]/[m$($^{82}$Se)
$\it-$ $\it m_e]$, is plotted using the mass values for $^{82}$Kr and
$^{82}$Se from the atomic mass evaluation AME2003 \cite{RefAudi2003}.

\begin{figure}[t]
\begin{center}
\includegraphics[scale=0.306,clip=]{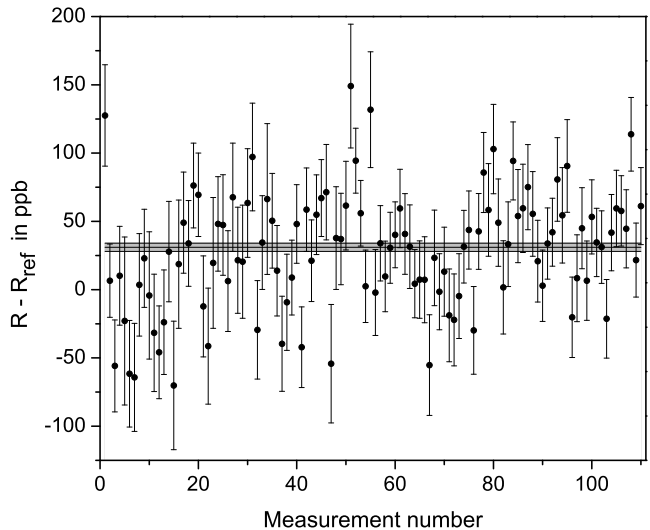}
\end{center}
\caption{Difference between the cyclotron frequency ratio of $^{82}$Kr$^{+}$
and $^{82}$Se$^{+}$ and the ratio obtained from literature mass data
\cite{RefAudi2003}. The solid lines indicate the weighted average and the
1$\sigma$ statistical uncertainty band.\label{Fig:2}}
\end{figure}

In preparation for, and during the measurement process, great care was taken
to minimize possible systematic effects. By measuring mass doublets,
contributions to the measurement uncertainty arising from mass dependent
systematic effects due to frequency shifts, for example caused by field
imperfections, are already greatly reduced. Nevertheless, prior to the
measurements, imperfections were carefully minimized following a tuning
procedure described in Ref.\ \cite{RefBrodeur2012}.

The effect of a non-linear magnetic field drift is not accounted for in the
data evaluation and is not mitigated by using a mass doublet. Therefore the
cyclotron frequency measurements of $^{82}$Kr$^{+}$ and $^{82}$Se$^{+}$ were
alternated with a period of no greater than 1 hour. Based on an earlier study
\cite{RefRingle2007}, this should lead to residual systematic effects of the
cyclotron frequency ratio no larger than 1 ppb. This uncertainty was further
minimized by stabilizing the pressure in the liquid helium dewar to 10 ppm
(parts per million) \cite{RefRingle2009}, resulting in an uncertainty well
below 1 ppb. Simultaneously trapped ions of a different $\it{m/q}$ value
\cite{RefBollen1992} can cause frequency shifts --- this effect was minimized
by verifying that contaminant ions were never present at a level exceeding a
few percent, low enough to not lead to a significant shift at the desired
precision.

The weighted average of the ratios for the individual runs and the
corresponding statistical uncertainty are listed in Table I together with the
weighted average $R_{\mathrm{LEBIT}}$ of all results and the value
$R_{\mathrm{AME2003}}$ obtained with mass data from AME2003
\cite{RefAudi2003}. Through evaluation of the entire data set, with
statistical errors as obtained from the fits of theoretical lineshapes to the
measured cyclotron resonance curves, we determined a Birge ratio
\cite{RefBirge1932} of 1.27(5). While close to unity, the significant
deviation indicates the presence of residual systematic effects at the 0.8 ppb
level not discovered in the individual measurements or in the tests for
systematic effects performed. Therefore, to account for these non-statistical
contributions we multiply the statistical uncertainty af the weighted average
$R_{\mathrm{LEBIT}}$ for all data by the value of the Birge ratio. Both the
statistical and total uncertainty for $R_{\mathrm{LEBIT}}$ are given in the
table.

 \begin{table}
\caption{\label{TableI}Average cyclotron frequency ratios
$R_{run}$$=\it{\nu^{int}_{c}}$($^{82}$Kr$^{+}$)$\it/\nu_c$($^{82}$Se$^{+}$)
with their statistical errors as obtained in four separate runs with N
frequency ratio measurements performed in each run. Also given is the final
weighted average $R_{\mathrm{LEBIT}}$ with its statistical and final
uncertainty and the ratio based on the AME2003 atomic mass evaluation
\cite{RefAudi2003}.}
 \begin{ruledtabular}
 \begin{tabular}{c c c}
Run & N & $R_{run}$\\
\hline
1 & 53 & \hspace{1 pt} 1.000 039 285(5)\\
2 & 2 & 1.000 039 30(2)\\
3 & 7 & 1.000 039 29(1)\\
4 & 48 &  \hspace{1 pt} 1.000 039 290(4)\\
\hline
$R_{\mathrm{LEBIT}}$ & &  \hspace{13 pt} 1.000 039 290(4)(5)\\
$R_{\mathrm{AME2003}}$ & & 1.000 039 26(3)\\
\end{tabular}
\end{ruledtabular}
\end{table}

The $\bb$ $Q$-value is determined from the mass difference between the mother
nuclide of mass $\it{m_m}$ and daughter nuclide of mass $\it{m_d}$ through:
\begin{equation} Q_{\beta\beta}/c^2 = m_m-m_d = \left (R-1\right )(m_d-m_e),
\end{equation} where $R$ is the cyclotron frequency ratio between the singly
charged ions of the daughter and mother nuclides, $c$ is the speed of light,
and $\it{m_{e}}$ accounts for the missing electron mass of singly charged ions
used in the measurement. Using our final frequency ratio $R_{\mathrm{LEBIT}}$
and the AME2003 mass for $^{82}$Kr we obtain, $\it{Q_{\beta\beta}}$ = 2
997.9(3) keV. The new LEBIT $Q$-value is nearly an order of magnitude more
precise than the previous value of $\it{Q_{\beta\beta}}$ = 2 996(2) keV based
on mass data from \cite{RefAudi2003}, a dramatic improvement to one of the
ingredients needed for a better determination of the half-life limit for
$\bbz$ in $^{82}$Se$^{+}$.


In addition to a precise $\qbb$-value, an accurate NME for the decay is also
needed. A number of methods have been applied to the NME problem, and in
$^{82}$Se one of the most prominent methods --- the shell model \cite{SM,SM2}
--- gives matrix elements that are less than half the size of those produced
by the quasiparticle random phase approximation (QRPA) \cite{QRPA}, the
Interacting Boson Model \cite{IBM}, and the Generator Coordinate Method
\cite{GCM}. Each approach has its strengths and weaknesses; for instance, the
shell model incorporates complicated correlations but only in a small
single-particle space (the valence shell), while the QRPA is applied in large
single-particle spaces but includes only simple correlations. Here we report
an attempt to correct for the shell model deficiency by calculating an effective $\bbz$
operator that implicitly includes effects of single-particle levels that are
outside the valence shell.

Our calculation, which will be described in detail in a forthcoming
publication \cite{EngelHolt}, uses diagrammatic perturbation theory to
construct an effective two-body $\bbz$ operator that, together with an
effective Hamiltonian, allows the shell model to get around its truncations
when the procedure is carried to completion \cite{LNP,Gmatrix}. This framework
has been used extensively to determine effective valence-space interactions,
but much less so to construct other effective operators. The only existing
work on an effective $\bbz$ operator is an exploratory calculation, as it
happens, for $^{82}$Se~\cite{EngelHagen}. In that work the starting point was
a $G$-matrix interaction~\cite{Gmatrix}. Using a low-momentum
interaction~\cite{Vlowk}, $\vlowk$, derived from nuclear forces in chiral
effective field theory~\cite{chiral}, we have carried this calculation
significantly further. We now include all diagrams to second order in the
interaction, state norms, and folds, while expanding the set of high-lying
single-particle orbits that we treat.

\begin{table}
 \caption{\label{NME} $^{82}$Se $\bbz$ matrix elements calculated in QRPA,
 standard shell model (SM), and with the effective $\bbz$ 
 operator discussed in the text.}
 \vspace{2mm}
 \begin{ruledtabular}
 \begin{tabular}{c c c}
QRPA~\cite{QRPA} & SM~\cite{SM2}  &  
 Corrected SM\\[1mm]
\hline
5.19 & 2.64 &  3.56 \\[1mm]
\end{tabular}
\end{ruledtabular}
\end{table}

Our result is shown in Table~\ref{NME}, which we compare to values obtained
from QRPA and the standard shell model. The contributions from outside the
valence space to the effective operator increase the shell model NME by $\sim
30 \%$ to 3.56, narrowing the gap between it and the QRPA.

This calculation represents an important first step in producing a true
\textit{ab initio} NME, suggesting that it will be larger than the shell model
has indicated. We plan to improve our results by including three-nucleon
forces~\cite{Oxygen,Calcium,RefGallant2012} and two-body currents~\cite{Menendez}, by
constructing an effective interaction consistent with the effective decay
operator, and by investigating the size of induced three-body terms in the
effective decay operator.


In conclusion, by using Penning trap mass spectrometry we have performed the
first direct $Q$-value measurement of $^{82}$Se $\bb$ by measuring the
cyclotron frequency ratio between singly charged ions of $^{82}$Se and the
$\bb$ daughter, $^{82}$Kr. Our result, $\it{Q_{\beta\beta}}$ = 2 997.9(3) keV,
is nearly an order of magnitude more precise than the previous value based on
the 2003 atomic mass evaluation \cite{RefAudi2003}. Following the procedure in
Ref.\ \cite{RefSuhonen1998} and using our new $Q$-value, we calculate the PSF
for the $\bbz$ mode of $^{82}$Se to be $G_{0\nu}$ = 2.848(1) x 10$^{-14}$
yr$^{-1}$, where the uncertainty has also been improved by nearly an order of
magnitude. With the corrected shell model NME calculation presented here and
the current upper limits of $\langle\it{m_{\beta\beta}}\rangle$ = 140 - 380
meV from the EXO-200 experiment \cite{RefEXO2012} we obtain a lower limit
range for the $^{82}$Se $\bbz$ half-life of 5.0 x 10$^{24}$ - 3.7 x 10$^{25}$
years. Assuming SuperNEMO achieves its projected sensitivity at the 90\%
confidence level of 1-2 x 10$^{26}$ years, an effective neutrino mass as low
as 60-85 meV could be detected.

We wish to acknowledge the support of Michigan State University, the National
Science Foundation under Cooperative Agreement No. PHY-11-02511, and the US
Department of Energy under Contracts DE-FG02-97ER41019, DE-FC02-07ER41457
(UNEDF SciDAC Collaboration), and DE-FG02-96ER40963 (UT).

\end{document}